\documentclass[11pt,a4paper]{article}
\usepackage{jheppub}
\usepackage[T1]{fontenc}
\usepackage{amssymb}
\usepackage{mathtools}
\usepackage{stackengine}
\usepackage{scalerel}

\usepackage{hyperref}
\usepackage{epsfig}
\usepackage{amsmath,latexsym,amssymb}
\usepackage{graphicx}
\usepackage{braket}
\usepackage{pdfsync}

\usepackage{framed}
\usepackage{mathtools}

\usepackage{color}
\usepackage{pstricks}

\usepackage{jheppub}
\usepackage[T1]{fontenc}
\usepackage{amssymb}
\usepackage{mathtools}
\usepackage{amsmath}
\usepackage{bm}
\usepackage{stackengine}

\usepackage{scalerel}
\usepackage{eufrak}
\usepackage{framed}

\usepackage{mathtools}

\usepackage{amsmath}
\usepackage{bbold}
\usepackage{bm}
\usepackage{latexsym}
\usepackage{braket}
\usepackage{slashed}
\usepackage{graphicx,booktabs,multirow}
\usepackage{eufrak}
\numberwithin{equation}{section}

\makeatletter
\newcommand{\doublewidetilde}[1]{{%
  \mathpalette\double@widetilde{#1}%
}}
\newcommand{\double@widetilde}[2]{%
  \sbox\z@{$\m@th#1\widetilde{#2}$}%
  \ht\z@=.9\ht\z@
  \widetilde{\box\z@}%
}
\makeatother

\usepackage{hyperref}
\usepackage{slashed}
\usepackage{epsfig}
\usepackage{amsmath,latexsym,amssymb}
\usepackage{graphicx}
\usepackage[latin1]{inputenc}
\usepackage{braket}
\usepackage{pdfsync}

\usepackage{framed}
\renewcommand{\L}{{\bm{L}}}
\renewcommand{\P}{{\bm{P}}}
\newcommand{\Q}{{\bm{Q}}}
\usepackage{mathtools}

\def\be{\begin{equation}}
\def\ee{\end{equation}}
\def\ba{\begin{eqnarray}}
\def\ea{\end{eqnarray}}

\def\ket#1{\mathinner{|{#1}\rangle}}

\def\braket#1{\mathinner{\langle{#1}\rangle}}

\renewcommand{\braket}[2]{\left< #1 \vphantom{#2} \right|
 \left. #2 \vphantom{#1} \right>} 
\newcommand{\matrixel}[3]{\left< #1 \vphantom{#2#3} \right|
 #2 \left| #3 \vphantom{#1#2} \right>} 

\newcommand{\comment}[1]{}

\newcommand{\eea}{\end{eqnarray}}

\author{
Tomasz R.\ Taylor${}^{1,2}$,\, Bin Zhu${}^{3}$\\[0.5cm]
 $^1${\it Department of Physics,
  Northeastern University, Boston, MA 02115, USA}\\
  $^2${\it Faculty of Physics, University of Warsaw, Pasteura 5, 02-093 Warsaw, Poland}\\
$^3${\it School of Mathematics and Maxwell Institute for Mathematical Sciences, University of Edinburgh,
EH9 3FD, UK }\\[0.2cm]
}

\emailAdd{taylor@neu.edu}
\emailAdd{bzhu@ed.ac.uk}

\title{\boldmath {Scattering of Quantum Particles in Global de Sitter Spacetime~II: Scalars in Deep Infrared} \unboldmath}

\abstract{We apply the S-matrix formalism developed in Part I to the interacting scalar theory in  four-dimensional de Sitter spacetime. The amplitudes are computed in the angular momentum basis, appropriate to the representations of $SO(1,4)$ de Sitter symmetry group. We discuss the properties of wavefunctions in Bunch-Davies vacuum and derive a new integral representation for the Feynman propagator. We focus on deep infrared processes probing the  large scale structure of spacetime, in particular on the processes that are normally forbidden by the energy-momentum conservation laws in flat spacetime. We find that there are no stable particles in self-interacting scalar field theory, but the decay rates are exponentially suppressed for particles with masses far above $\hbar/c\ell$, where $\ell$ is the de Sitter radius. We also show that the ``all out'' amplitudes describing multiparticle production from the vacuum are identically zero, hence Bunch-Davies vacuum is stable with respect to the matter interactions. We show that at the tree level, all scattering amplitudes are infrared finite, well-defined functions of quantum numbers. They have no kinematic singularities, except for the processes involving conformally coupled scalars.}

\gdef\@fpheader{}
\makeatother

\begin{document}
\maketitle
\section{Introduction}

In this work, we investigate the scattering processes of elementary particles in maximally symmetric de Sitter spacetime. 
Our approach is centered on the observers and on the observables measured along their worldlines, as the proper time elapses in inertial reference frames idling on timelike geodesics. The {\em global} maximal de Sitter symmetry is absolutely crucial because it allows studying all scattering events in one particular reference frame, by using it in exactly the same way as Poincar\'e symmetry is used in Minkowski spacetime in formulating relativistically invariant Quantum Field Theory. 

With de Sitter symmetry adopted as the guiding principle in Part I \cite{Taylor:2024vdc}, we considered  the scattering processes of particles belonging to the representations of de Sitter group. We explained how the states that probe asymptotically short invariant distances  morph into the representations of Poincar\'e group. At the level of symmetry algebras, this corresponds to the so-called In\"on\"u-Wigner contraction \cite{ew}.

The $d$-dimensional de Sitter manifold dS$_d$ can be constructed  by imposing a hyperbolic constraint in $(d+1)$-dimensional ``embedding'' Minkowski spacetime. de Sitter symmetry appears then as the $SO(1,d)$ Lorentz symmetry of the embedding space. 
The embedding formalism allows identifying the time evolution operator of free quantum states as the boost generator. Hence in the interacting theory, the corresponding Hamiltonian, describing quantum particles interacting in observer's frame, is associated to the Killing boost vector.

The main result of Part I \cite{Taylor:2024vdc} is a simple formula for the scattering amplitude describing the transition from an incoming state 
 $\ket{\alpha}^{in}\equiv \ket{\alpha(\tau=-\infty)}$,  observed as a state of  free particles  in the infinite past (observer's proper time $\tau\rightarrow-\infty$),  to an  outgoing state $\ket{\beta}^{out}\equiv\ket{\beta(\tau=+\infty)}$, observed  in the infinite future (observer's $\tau\rightarrow+\infty$): 
\be ^{out}\braket{\beta\,}{\!\alpha}^{in}
=\matrixel{\beta(0)}{T\, \makebox{exp}\Big(-i\int d^dx
\sqrt{-g}\,H_I[\phi(x)]
\Big)}{\alpha(0)}\ ,\label{dys}\ee
where $H_I$ denotes the interaction Hamiltonian density while $\phi(x)$ represent generic quantum fields. The time ordering $T$ is with respect to  the canonical time coordinate of the embedding Minkowski spacetime, which runs in the same direction as the proper time on observer's worldline.  Eq.(\ref{dys}) extends the validity of well-known Dyson's formula to de Sitter spacetime. The amplitudes do not depend on the coordinates used to parametrize de Sitter spacetime, as long as they preserve canonical time ordering.

We should mention two earlier works, Refs.\cite{Parikh:2002py} and \cite{ Marolf:2012kh}, that influenced our construction. In Part I, we listed more references and compared our approach with the existing literature.

The paper is organized as follows. In Section 2, we discuss quantum scalar fields in four-dimensional de Sitter spacetime. The wave functions are written as linear combinations  of hyperspherical harmonics on $S_3$ weighted by time-dependent associated Legendre functions. We identify the positive and negative frequency modes for the observer remaining for a short proper time near the North pole. In  this neighborhood, plane waves with short wavelength can be constructed as a superposition of hyperspherical harmonics and for a short time, local physics seems to be indistinguishable from physics in flat spacetime. This is the time when  the observer performs scattering experiments and measures transition amplitudes with the intention of probing larger distances. We quantize scalar fields and identify the vacuum as de Sitter invariant Bunch-Davies state \cite{Bunch:1978yq}. In Section 3, we discuss the Feynman propagator and derive a new integral representation which is suitable for Feynman diagram computations in the angular momentum basis. While in Minkowski spacetime, the propagator propagates virtual particles with definite (off-shell) four-momentum, de Sitter propagator propagates virtual wavepackets with a wide spread of (angular) momenta. In Section 4, we discuss some general properties of tree-level scattering amplitudes, including infrared finiteness and conservation laws. We then proceed to the computations of transition amplitudes for deep infrared processes probing large distances, comparable to de Sitter radius. These processes involve light particles and small (angular) momenta. We focus on the processes forbidden by the energy-momentum conservation laws in flat spacetime. We conclude in section 5. In the Appendix, we collect some formulas applied in the paper; they describe the properties of associated Legendre functions and Gegenbauer polynomials entering into the wavefunctions.
\section{Scalars in $d=4$}
We begin by quantizing free scalar field theory. We will expand on the foundational work of Chernikov and Tagirov \cite{Chernikov:1968zm}, and Mottola \cite{Mottola:1984ar}, and organize the material in a  way similar  to $d=2$  presented in Part I. 
\subsection{Coordinates, metrics and notation}
Four-dimensional de Sitter manifold dS has the topology of $\mathbb{R}^1\times S_3$.   It can be realized as a hypersurface described by the equation
\be -X_0^2+X^2_1+\dots +X^2_4=\ell^2\ee
in the embedding $d=5$ Minkowski space \cite{Spradlin:2001pw}.
The $SO(1,4)$ de Sitter isometry group follows from Lorentz symmetry of the constraint.
The parameter $\ell$ with units of length is called de Sitter radius.
It is related to the  curvature scalar and the related cosmological constant in the following way:
\be R={12\over \ell^2}\ ,\qquad
\Lambda={3\over \ell^2} \, .\ee
We will be using  global (conformal) coordinates $(t,\chi,\theta,\phi)$. The spatial $S_3$ are parameterized by the radii that depend on the conformal time coordinate $t\in [-\pi/2,\pi/2]$, two angular polar coordinates $\chi\in [0,\pi], \theta\in[0,\pi]$ and one azimuthal angular coordinate $\varphi\in [0,2\pi]$. We will be often using one symbol,  $\Omega=(\chi,\theta, \varphi)$, to collectively denote the angular coordinates. The global coordinates are related to the embedding ones in the following way:
\begin{align}
X^0&= \ell\tan t\ \equiv\ T\, , \nonumber\\ X^1 &= \frac{\ell}{\cos t}\cos \chi \, , \nonumber\\ X^2 &= \frac{\ell}{\cos t} \sin\chi \cos\theta \, , \\
X^3 &= \frac{\ell}{\cos t} \sin\chi \sin\theta \cos\varphi \, ,\nonumber \\ X^4 &= \frac{\ell}{\cos t} \sin\chi \sin\theta \sin \varphi \, . \label{eq:embeddingcoor}\nonumber
\end{align}
In terms of global coordinates, the pullback of Lorentz metrics on de Sitter manifold read
\begin{equation}
ds^2 = \frac{\ell^2}{\cos^2 t} \left( -dt^2 + d\Omega^2\,\right),\label{eq:dSmetric}
\end{equation}
where
\begin{equation}
d\Omega^2 =  d\chi^2 +\sin^2\chi (d\theta^2+\sin^2\theta \,d\varphi^2)\label{msph}
\end{equation}
is the metric on  unit $S^3$. From now on, we set de Sitter radius $\ell=1$. Then the embedding time coordinate $T=\tan t$.
\subsection{Scalar wave equation and hyperspherical harmonics}
The Klein-Gordon wave equation for the scalar field $\phi(t,\Omega)$ with the Lagrangian mass parameter $m$ reads
\begin{equation}
(\square -m^2) \phi(t,\Omega) = 0 \, . \label{eq:waveeq}
\end{equation}
In global coordinates, the d'Alembertian is given by
\begin{align}
\square = - \cos^2t \left( \frac{\partial^2}{\partial t^2} + 2 \tan t \frac{\partial}{\partial t}\right) +\cos^2 t \, \Delta_{S_3} \, ,
\end{align}
where $\Delta_{S_3}$ is the Laplace operator on $S_3$. 

The eigenfunctions of $\Delta_{S^3}$ are known as the hyperspherical $SO(4)$ harmonics \cite{Avery:1989} $Y_{\bm{L}}(\Omega)\equiv Y_{Lln}(\chi,\theta,\varphi)$ and satisfy
\begin{equation}
\Delta_{S_3} Y_{L l n}(\chi, \theta, \varphi) = -L(L+2) Y_{L l n}(\chi, \theta, \varphi) \, . \
\end{equation}
They are labelled by integers $L,l,n$, subject to the constraints 
\begin{equation} L\ge 0,\qquad l=0,1,\dots, L,\qquad n=-l,-l+1,\dots,l-1, l.\ee
We will often use $\bm{L}=(L,l,n)$ to collectively denote these indices. Hyperspherical harmonics can be expressed in terms of the standard spherical harmonics $Y_{ln}(\theta,\varphi)$ in the following way:
\begin{equation}
Y_{\bm{L}} (\Omega) = N_{\bm{L}\,} \sin^l\chi\, C_{L-l}^{1+l}(\cos\chi)\, Y_{ln}(\theta,\varphi) \, , \label{eq:Avery1}
\end{equation}
 where $C_{L-l}^{1+l}(\cos\chi)$ are
the Gegenbauer polynominals. See Appendix \ref{app:A} for their definition and basic properties. These functions are normalized with respect to the $S_3$ metrics (\ref{msph}), with
\begin{equation}
N_{\bm{L}} = \left( \frac{2}{\pi} \frac{(2l)!! (L+1) (L-l)! (2l+1)!}{(2l+1)!! (L+l+1)!} \right)^{\frac{1}{2}} \, . \label{eq:Avery2}
\end{equation}
\subsection{Wavefunctions}
In order to solve the wave equation, we decompose the scalar field into hyperspherical harmonics:
\be \phi(t,\Omega)=\sum_{\bm{L}}f_L(t)Y_{\bm{L}}(\Omega)\label{wavf}\ee
where the sum
\be \sum_{\bm{L}}\equiv \sum_{L=0}^\infty\,\sum_{l=0}^{L}\,\sum_{n=-l}^{l}.
\ee
Then the wave equation (\ref{eq:waveeq}) implies the following differential equations for the time-dependent expansion coefficients
\begin{equation}
\Big[ \cos^2t \left( \frac{d^2}{d t^2} + 2 \tan t \frac{d}{d t}\right)+L(L+2) \cos^2 t  +m^2\Big]f_L(t) = 0 \, . \label{eq:diffft}
\end{equation}
By following the same steps as in the case of two dimensions \cite{Taylor:2024vdc}, Eqs.(\ref{eq:diffft}) can be reduced to Legendre equations. The general solution is a linear combination of two functions:
\begin{align}
f_{1L}(t) &= \sqrt{ \cos t\,}^3
 P^{-i\mu}_{L+\frac{1}{2}}(\sin t) \, ,\nonumber \\
f_{2L}(t) &= \sqrt{ \cos t\,}^3Q^{-i\mu}_{L+\frac{1}{2}}(\sin t) \, ,\label{lsol}
\end{align}
where $P$ and $Q$ are the associated Legendre functions of the first and second kind, respectively. The parameter $\mu$, which enters into the order (defined below) of these functions, is given by
\begin{equation}
\mu= \sqrt{m^2-\frac{9}{4}} \label{eq:msquare}.\ee
Note that $m=3/2$ is the ``critical'' value at which the order turns from real to imaginary. This value separates two different $SO(1,4)$ representations of the respective spin zero quanta, the principal and complementary series \cite{rep1,rep2,rep3,rep4}. At this point, to be specific,  we assume real $\mu$, {\it i.e}.\ $m^2\ge 9/4$.

The associated Legendre functions will appear all through this work. They are defined in Appendix \ref{app:A}. We will be using the following notation to distinguish various types of functions. The functions $\P^{\mu}_{\nu}(z)$ and $\Q^{\mu}_{\nu}(z)$ ($\mu$ is  the order and $\nu$ is the degree) are defined in terms of hypergeometric and elementary functions. The argument $z$ lies in the complex plane with a cut extending from $-\infty$  to $+1$. These  functions have highly nontrivial monodromy properties, therefore their analytic continuation requires special care. On the other hand, the argument of the functions $P$ and $Q$ entering the solutions (\ref{lsol}), $\sin t\in [-1,1]$, is situated on the cut, therefore the function values depend on the way the argument approaches the cut. These functions, also known as Ferrers' functions, will be denoted by $P^{\mu}_{\nu}(x)$ and $Q^{\mu}_{\nu}(x)$ and have the arguments $x\in [-1,1]$. They are defined in Appendix \ref{app:A} by taking certain combinations of $\P^{\mu}_{\nu}(z)$ and $\Q^{\mu}_{\nu}(z)$ with $z$ approching $x$ from the regions below and above the cut ($z\to x\mp i\epsilon)$.

\subsection{Positive and negative frequencies}
The quantization of scalar fields in flat spacetime begins by identifying the positive and negative frequency modes of the wavefunctions. The positive frequency modes are then identified as the wavefunctions of particles created by the fields. Their complex conjugate, negative frequency modes are often interpreted as the wavefunctions of (anti)particles propagating backward in time. de Sitter wavefunctions (\ref{lsol}) contain, however, entire spectrum of positive and negative frequencies, with the relative weights depending on
reference frames. Since all geodesic observers are related by de Sitter symmetry transformations, we can always choose the observer located at the North pole $\chi=0$ with the proper time  reset to $\tau=0$
at $T=0$. Then
\be \tau=\sinh^{-1}T=\ln\sqrt{\frac{1+\sin t}{1-\sin t}}\ , \qquad\chi(\tau)=0\ .\ee
As in Part I, in order to construct the  wavefunctions analogous to positive frequency modes in Minkowski space, we consider the waves that probe
spacetime at asymptotically short invariant distances. The wavelengths of spherical waves described by the harmonics $Y_\L(\Omega)$ become short in the limit of
$\L\to\infty$, {\em i.e}.\ when all integers $ |n|\to \infty,~l\to\infty,~L\to\infty$, ordered as $|n|\ll l\ll L$. We require that in observer's neighborhood, such waves propagate with positive frequencies. As shown in Part I, up to an overall normalization factor, there is a unique combination of large degree $L+\frac{1}{2}$ Ferrers' functions that contains only positive frequencies:
\begin{align}
f_{L}(t)_+ &=\sqrt{\cos t\,}^3\, \big[P^{- i\mu}_{L+\frac{1}{2}}(\sin t) - \frac{2i}{\pi} Q^{- i\mu}_{L+\frac{1}{2}}(\sin t)\big]\label{pw}\\[1mm] &~~~~~\sim\, e^{- i (L+1)t}\cos t \,\sqrt{\frac{2}{\pi}} e^{
\mu\pi\over 2}\left(L+\frac{1}{2}\right)^{- i\mu-\frac{1}{2}} e^{ i\frac{\pi}{2} (L+\frac{1}{2})} \, .\nonumber
 \end{align}
This large $L$ asymptotic expansion is valid up to ${\cal O}(1/L)$ corrections, and only when $t\in [-\pi/2+1/L, \pi/2-1/L]$, {\it i.e}.\ it is not a good approximation in the far past or far future  at large $|\tau|$.  We will examine this asymptotic behavious in more detail later, by using different representations of wavefunctions.

We will be also using two alternative representations of wavefunctions. The first one follows from the definition (\ref{eq:defQcut}) of Ferrers' functions:
 \be
 P^{- i\mu}_{L+\frac{1}{2}}(\sin t) - \frac{2i}{\pi} Q^{- i\mu}_{L+\frac{1}{2}}(\sin t)=\frac{2e^{-\frac{\mu\pi}{2}}}{i\pi}\Q^{- i\mu}_{L+\frac{1}{2}}(\sin t-i\epsilon)\ .\ee
The second one can be obtained by combining the reflection formula (\ref{eq:refformula}) for the $\Q$ function with the Whipple's formula (\ref{eq:Whipple}):
 \be \Q^{- i\mu}_{L+\frac{1}{2}}(\sin t-i\epsilon)=e^{i\pi\over 4}e^{\mu\pi}\Gamma(L+\frac{3}{2}-i\mu)\sqrt{\pi\over 2\cos t}\P^{-L-1}_{-\frac{1}{2}+i\mu}(i\tan t)\times\left\{ 
{1~~~~~~~~~~~(t>0)\atop (-1)^{L+1}~~(t<0)}
\right.\label{meh}\ee
The Legendre functions $\P$ of degree $\nu=-\frac{1}{2}+i\mu$, encountered in the above relation, are known as Mehler's or conical functions. Note that the argument $i\tan t=iT$ is the Wick-rotated time coordinate of the embedding Minkowski space.
As pointed out by Mottola in Ref.\cite{Mottola:1984ar}, these functions can be obtained from $d=5$ [$SO(5)$] Euclidean hyperspherical harmonics by a Wick rotation. They describe scalar particles in the  so-called Bunch-Davis a.k.a.\ Euclidean vacuum
\cite{Bunch:1978yq}. The factor $(-1)^{L+1}$ on the r.h.s.\ of Eq.(\ref{meh}) compensates for the discontinuity of Mehler's function across the cut on the real axis. The continuity of l.h.s.\ is explicit for all $t$ in the interval $ [-\pi/2, \pi/2]$.
\subsection{Hyperspherical harmonics and plane waves}
The wavefunctions (\ref{wavf}) factorize into time-dependent Legendre functions and hyperspherical harmonics, which 
 are the eigenfunctions of quadratic Casimir operators of $SO(4)$, $SO(3)$ and $SO(2)$ subgroups of the $SO(1,4)$ de Sitter symmetry group. The corresponding quantum numbers, $L, ~l$ and $n$, respectively, label the states belonging to de Sitter symmetry multiplets. 
Hence it is appropriate to call (\ref{wavf}) the wavefunctions in the angular momentum basis. The observer may prefer, however, to use wavefunctions similar to the linear momentum basis (plane waves) in her/his/their neighborhood, where spacetime appears to be flat. For the observer on the North pole, it is the region of small $\chi$. Note that $\chi$ is a radial coordinate in tangent space. Below, we explain how plane waves with short wavelengths (large spatial momentum $\bm{k}$) can be constructed from spherical waves with large angular momentum $\bm{L}$. Indeed, the In\"on\"u-Wigner contraction of de Sitter to Poincar\'e symmetry algebras, described in Part I, yields $|\bm{k}|\approx L$ for large $L$, in the $\ell\to \infty$ flat limit.

It is well known that three-dimensional plane waves can be expanded into spherical harmonics:
\be
e^{i\bm{k}\cdot \bm{r}} = \sum_{l=0}^\infty \sum_{n=-l}^l 4\pi i^l j_l(kr) Y_{ln}(\hat{r}) Y_{ln}^*(\hat{k})\ ,
\ee
where $\hat{r}$ and $\hat{k}$ are the angles specifying the  directions of $\bm{r}$ and $\bm{k}$, respectively, while $r$ and $k$ are their magnitudes.
The coefficients $j_l$ are the spherical Bessel functions of the first kind. In observer's neighborhood, we identify $r=\chi$, $\hat{r}=(\theta,\phi)$ and consider small $\chi$. To make a connection with  hyperspherical harmonics (\ref{eq:Avery1}), we take the large $L\gg l$ limit by using the asymptotic formula \cite{Durand} for Gegenbauer functions with large parameters:
\begin{equation}
C_{L-l}^{1+l}(\cos \chi) = \frac{1}{\Gamma(1+l)} \Big( {L+1\over 2\sin \chi}\Big)^{l+\frac{1}{2}}\sqrt{2y}\,j_l(y)~ +~ {\cal O}(L^{-2/3}),\ee
where
\begin{equation}
y= \sqrt{2(L+1)^2(1-\cos \chi)} \approx L\chi\ .
\end{equation}
Taking into account the normalization factors (\ref{eq:Avery2}),  we find that in the large $L$ limit, for small $\chi$:
\be
Y_{Lln}(\chi, \theta, \varphi)=L\sqrt{2\over\pi} j_l(L\chi) Y_{ln}(\theta,\varphi)\ ,
\ee
therefore
\be
e^{i\bm{k}\cdot \bm{r}} = \frac{(2\pi)^{3/2}}{L}\sum_{l=0}^\infty \sum_{n=-l}^l i^l  Y_{ln}^*(\hat{k})\,Y_{Lln}(\chi,\theta,\varphi)\ ,
\label{rhar}\ee
with
\be r=\chi, \qquad \hat{r}=(\theta,\phi),\qquad |\bm{k}|=L.\label{rid}\ee
\subsection{Quantization}
We conclude that the scalar field can be expanded as
\be \phi(t,\Omega)=\sum_{\bm{L}}\Big(a_{\bm{L}}\phi_{ \bm{L}}(t,\Omega)_{+}+a^\dagger_{\bm{L}}\phi_{ \bm{L}}(t,\Omega)_{-}\Big)\ ,
\label{fiel}\ee
where $\phi_{ \bm{L}}(t,\Omega)_{+}$ are the positive frequency wavefuctions, normalized with respect to the Klein-Gordon norm and $\phi_{ \bm{L}}(t,\Omega)_{-}$ are their complex conjugates.
With the time-dependent factor expressed in terms of the conical functions (\ref{meh}), they are given by
\begin{align}
\phi_{ \bm{L}}(t,\Omega)_+ &= \frac{|\Gamma(L+\frac{3}{2}+i\mu)|}{\sqrt 2}\cos t\, \widetilde{\P}^{-L-1}_{-\frac{1}{2}+i\mu}(i \tan t) \, Y_{\bm{L}}(\Omega)\ ,\nonumber \\[1mm]
\phi_{ \bm{L}}(t,\Omega)_- &= \frac{|\Gamma(L+\frac{3}{2}+i\mu)|}{\sqrt 2}\cos t \,\widetilde {\P}^{-L-1}_{-\frac{1}{2}+i\mu}(-i \tan t) \, Y_{\bm{L}}^*(\Omega)\ ,  \label{eq:phi4d}
\end{align}
where
\be \widetilde{\P}^{-L-1}_{-\frac{1}{2}+i\mu}(\pm i\tan t) =    {\P}^{-L-1}_{-\frac{1}{2}+i\mu}(\pm i\tan t) \times    \left\{ 
{(\mp i)^{L+1}~~~~~(t>0)\atop(\pm  i)^{L+1}~~~~~(t<0)}
\right.\label{meh1}\ee
Note that the phases of normalization factors are chosen in such a way that 
\be \widetilde{\P}^{-L-1}_{-\frac{1}{2}+i\mu}(-i\tan t) =\widetilde{\P}^{-L-1}_{-\frac{1}{2}+i\mu}(i\tan t)^*.\label{meh22}\ee
They also ensure continuity at $t=0$.
Alternatively,
\begin{align}
\phi_{ \bm{L}}(t,\Omega)_+ &= \frac{e^{-\mu \pi}e^{-i\frac{\pi}{4}}}{\sqrt{\pi}} \sqrt{\Gamma(L+\frac{3}{2}+i\mu)\over \Gamma(L+\frac{3}{2}-i\mu)}
(-i)^{L+1}\sqrt{\cos t}^3
\Q^{-i\mu}_{L+\frac{1}{2}}(\sin t-i\epsilon) \, Y_{\bm{L}}(\Omega)\, ,
 \nonumber\\[1mm]
\phi_{ \bm{L}}(t,\Omega)_- &= \frac{e^{\mu \pi}e^{i\frac{\pi}{4}}}{\sqrt{\pi}} \sqrt{\Gamma(L+\frac{3}{2}-i\mu)\over \Gamma(L+\frac{3}{2}+i\mu)}
i^{L+1} \sqrt{\cos t}^3\Q^{i\mu}_{L+\frac{1}{2}}(\sin t+i\epsilon) \, Y_{\bm{L}}^*(\Omega)\, .  \label{phi4intermsofQ}
\end{align}
Furthermore,
\begin{align}
\phi(t,\Omega)_+ = \frac{e^{- \frac{\mu\pi}{2}} e^{-i\frac{\pi}{4}} \sqrt{\pi}}{2}&
\sqrt{\Gamma(L+\frac{3}{2}+i\mu)\over \Gamma(L+\frac{3}{2}-i\mu)}
(-i)^{L}
\nonumber \\ &
\times \sqrt{\cos t}^3 \big(P^{-i\mu}_{L+\frac{1}{2}}(\sin t) -\frac{2i}{\pi} Q^{-i\mu}_{L+\frac{1}{2}}(\sin t) \big) Y_{\bm{L}}(\Omega)\, ,\nonumber\\
\phi(t,\Omega)_-= \frac{e^{- \frac{\mu\pi}{2}} e^{i\frac{\pi}{4}}  \sqrt{\pi}}{2}&
 \sqrt{\Gamma(L+\frac{3}{2}-i\mu)\over \Gamma(L+\frac{3}{2}+i\mu)}\,
i^{L}\nonumber\\ & \times
\sqrt{\cos t}^3 \big(P^{+i\mu}_{L+\frac{1}{2}}(\sin t) +\frac{2i}{\pi} Q^{+i\mu}_{L+\frac{1}{2}}(\sin t) \big)Y_{\bm{L}}^*(\Omega) \, .\label{lwa}
\end{align}

In quantum theory, the creation and annihilation operators satisfy the commutation relations
\be[a_{ \bm{L}},a_{ \bm{L}'}^\dagger]=\delta_{{ \bm{L}}{ \bm{L}'}}\, ,\qquad [a_{ \bm{L}},a_{ \bm{L}'}]=[a_{ \bm{L}}^\dagger,a_{ \bm{L}'}^\dagger]=0\ .\label{crel}\ee
The vacuum state is annihilated by all annihilation operators:
\be a_{ \bm{L}}|0\rangle=0\ .\ee
This vacuum is {\it unique}, de Sitter invariant and common to all observers. Since it is related to Euclidean hyperspherical harmonics, it is usually  called  Euclidean, a.k.a.\ Bunch-Davies vacuum \cite{Mottola:1984ar,Bunch:1978yq}.  One-particle states are obtained by acting on this vacuum with the  creation operators:
\be |\bm{L},\mu\rangle =a_{ \bm{L}}^\dagger|0\rangle .\label{onep}\ee
{}For real $\mu$, they form the principal series representations of the $SO(1,4)$ de Sitter symmetry group with dimensions $\Delta=\frac{3}{2}+i\mu$ \cite{rep1,rep2,rep3,rep4}.
\section{Feynman propagator}
\subsection{Integral representation}
Our formulation of de Sitter QFT  follows the standard operator formulation of QFT in Minkowski spacetime. In this framework, the Feynman propagator of a free scalar field is given by
\be D_F(x,y)= \Theta(x^0-y^0)[\phi(x)_+,\phi(y)_-]+\Theta(y^0-x^0)[\phi(y)_+,\phi(x)_-]\ .\label{dpr}\ee
In de Sitter spacetime, this leads to
\begin{align}
D_F(t_1,\Omega_1;t_2,\Omega_2)=&~\Theta(t_1-t_2)\sum_{\bm{L}}
\phi_{ \bm{L}}(t_1,\Omega_1)_+\phi_{ \bm{L}}(t_2,\Omega_2)_-\nonumber \\ &
~+\Theta(t_2-t_1)\sum_{\bm{L}}
\phi_{ \bm{L}}(t_2,\Omega_2)_+\phi_{ \bm{L}}(t_1,\Omega_1)_-\ .\label{summ}
\end{align}

As a de Sitter scalar, the propagator  $D_F(x_1,x_2)$ must depend on the invariant distance, which is related to $(X_1-X_2)^2$  in the embedding space. It is convenient to define
\be z(x_1,x_2)=1-\frac{1}{2}(X_1-X_2)^2\ .\ee
Indeed, $z(x_1,x_2)$ is determined the geodesic distance $d(x_1,x_2)$ between two points:
\be  z(x_1,x_2) =\cos[d(x_1,x_2)]\ .\ee
It takes the following values depending on their separation:
\be z(x_1,x_2)\begin{cases} >1~~\makebox{timelike }\\ =1~~\makebox{lightlike} \\ <1~~\makebox{spacelike}\end{cases}\ee
First, let us consider the case of spacelike $z=x\in [-1,1]$. The endpoint $x=-1$ corresponds to $d(x_1,x_2)=\pi$, when the two points are antipodal ($X_1=-X_2$), while $x=1$ corresponds to a lightlike separation (including $X_1=X_2$).
From the Klein-Gordon equation, it follows that $D_F(x) $ can be written as $(1-x^2)^{-1/2}$ times a linear combination of $P^1_{-\frac{1}{2}+i\mu}(-x)$ and $Q^1_{-\frac{1}{2}+i\mu}(-x)$. 
In Euclidean vacuum, there is no singularity when two points are antipodal, therefore 
the second Ferrers' function $Q^1_{-\frac{1}{2}+i\mu}(-x)$,  which is singular at $x=-1$, is excluded from the solution.  On the other hand, the singularity at $x=1$ should match the light-cone/short-distance singularity of the flat space propagator:
\be
D_F(z)\to \frac{1}{4\pi^2(d^2+i\epsilon)} \quad \makebox{as}~d\to 0\ .\ee
This requirement leads to
\be
D_F(z)=\frac{1}{8\pi\cosh(\mu\pi)}\frac{\P^1_{-\frac{1}{2}+i\mu}(-z+i\epsilon)}{\sqrt{-1+(-z+i\epsilon)^2}}\ .       \label{pros}    \ee

Direct computation of dS propagators, by summing over the modes (\ref{summ}) instead of solving differential equations, is technically more difficult than in Minkowski space \cite{Fukuma:2013mx,Derezinski:2024ims}. It involves summing products of associated Legendre functions with nontrivial monodromies. A similar computation in Minkowski space, which amounts to integrating over the plane wave momentum modes, is relatively straightforward. It leads to the familiar momentum space representation of the propagator, in a form suitable for Feynman diagram computations. The propagator (\ref{pros}) has no such representation, but we can cast it in a form that will allow a direct comparison with the flat case. To that end, we will use the following integral representation:
\be
\widetilde{\P}^{-L-1}_{-\frac{1}{2}+i\mu}(i T) =\sqrt{2(1+T^2)^{L+1}\over\pi} \big|\Gamma(L+\frac{3}{2}+i\mu)\big|^{-2}  \int_0^\infty 
u^{L+\frac{1}{2}}K_{i\mu}(u)e^{-iuT}du\ ,
 \ee
which holds up to a constant phase factor. Here, $K_{i\mu}$ is the Bessel function of the second kind, of imaginary order $i\mu$. By using this representation in Eq.(\ref{dpr}) with $\tan t=T$, we obtain
\begin{align} D_F(T_1,\Omega_1; T_2,\Omega_2)=& \frac{1}{\pi}\sum_{\L}\Big\{\big[(1+T_1^2)(1+T^2_2)\big]^{L/2} \,\big|\Gamma(L+\frac{3}{2}+i\mu)\big|^{-2}  \,
Y_{\bm{L}}(\Omega_1)Y_{\bm{L}}^*(\Omega_2)\nonumber \\
&~~~\times\!\int_0^\infty \!\!\!\int_0^\infty \!\!dt\,du
\, (tu)^{L+\frac{1}{2}}K_{i\mu}(t)K_{i\mu}(u)\\[1mm] &~~~~~~~~~~~~\times\!\big[\Theta(T_1-T_2)e^{i(uT_2-tT_1)}+
\Theta(T_2-T_1)e^{i(uT_1-tT_2)}\big]\Big\} \, .
\nonumber\end{align}
The time-ordered factor can be written as
\begin{align} \Theta(T_1-T_2)e^{i(uT_2-tT_1)}~+~ &
\Theta(T_2-T_1)e^{i(uT_1-tT_2)} \\ &=\frac{i(t+u)}{2\pi}e^{i(T_1+T_2)(u-t)/2}\int_{-\infty}^{\infty}dk {e^{-ik(T_1-T_2)}\over k^2-(\frac{t+u}{2})^2+i\epsilon}
 \, .\nonumber\end{align}
In this way, we obtain
\begin{align} D_F(T_1,\Omega_1; &T_2,\Omega_2)=
\frac{i}{2\pi^2}\int_{-\infty}^{\infty}dk\sum_{\L}\Big\{e^{-ik(T_1-T_2)} Y_{\bm{L}}(\Omega_1)Y_{\bm{L}}^*(\Omega_2)\nonumber \\[1mm]
&~~~~~~\times \big[(1+T_1^2)(1+T^2_2)\big]^{L/2} \,\big|\Gamma(L+\frac{3}{2}+i\mu)\big|^{-2}  \,\label{despro}
 \\[1mm]
&\times\!\int_0^\infty \!\!\!\int_0^\infty \!\!dt\,du
\, (tu)^{L+\frac{1}{2}}(t+u)K_{i\mu}(t)K_{i\mu}(u){e^{i(T_1+T_2)(u-t)/2}\over k^2-(\frac{t+u}{2})^2+i\epsilon}\Big\} \, .\nonumber
\nonumber\end{align}

There is some similarity between Eq.(\ref{despro}) and the flat space propagator:
\be D_F(x,y)\big|_{{\cal M}_4}=\int {dk^0 d^3\bm{k}\over (2\pi)^4}\, e^{i k(x-y)}\,{i\over (k^0)^2-\bm{k}^2-m^2+i\epsilon} \, .\ee
The integral over three-momenta $\bm{k}$ is replaced in Eq.(\ref{despro}) by the sum over angular momenta $\L$. We will show  at the end of Section 3.2 that the sums and integrals are indeed equivalent in the region  of large $\L\sim \bm{k}$. The integral over the energies $k^0$ is replaced by the integral over the parameter $k$. The main difference is the presence of two additional integration parameters $t$ and $u$, with $[(t+u)/2]^2$  replacing $\bm{k}^2+m^2$ in the denominator of the flat space propagator. This is not surprising because in flat spacetime, the  momentum magnitude   $|\bm{k}|$ (wavelength)   determines the energy  $k^0$ (frequency) through the Lorentz-invariant dispersion relation, while in dS, a wavefunction with given $L$ contains a wide spectrum of frequencies.
\subsection{The limit of large mass and angular momentum}
In the limit of large mass $(\mu\to \infty)$ and large angular momentum ($L\to\infty $), the sum over  the modes (\ref{summ}) should be equivalent to the sum of plane wave modes, at least in the  neighborhood of the observer located at the North pole, for the waves propagating during a short time interval $(T_1, T_2)$ with $T_1^2\approx T_2^2\approx 0$. Here, we consider the limit
\be \mu\to\infty,\qquad L\to \infty,\qquad \alpha\equiv\frac{L}{\mu}~~\makebox{fixed.}\label{limes}\ee
of the integral
\be I(\mu,L)=\mu^{2(L+2)}\int_0^\infty \!\!\!\int_0^\infty \!\!dt\,du
\, (tu)^{L+\frac{1}{2}}(t+u)K_{i\mu}(\mu t)K_{i\mu}(\mu u){e^{i\mu(T_1+T_2)(u-t)/2}\over k^2-\mu^2(\frac{t+u}{2})^2+i\epsilon}\ ,\ee
which is the same integral as in the third line of Eq.(\ref{despro}), but with the  integration variables rescaled by $\mu$. For large $\mu$, the Bessel functions have ``uniformly valid'' asymptotic expansions \cite{bessel}
\be K_{i\mu}(\mu t)=\sqrt{\pi\over 2\mu}\frac{1}{(t^2-1)^{1/4}}\exp\Big[-\mu\xi(t)-{\pi\mu\over 2}+\dots\Big]\qquad (\makebox{as}~\mu\to \infty)\ ,\ee
where 
\be \xi(t)=\sqrt{t^2-1}-\tan^{-1}(\sqrt{t^2-1})\ .\ee
In the limit of large $L=\alpha\mu$, the integrands are suppressed by the exponential factor $e^{-\mu[f(t)+f^*(u)]}$,
with
\be
f(t)=\xi(t)-\alpha\ln t+itT\ ,\qquad T=\frac{T_1+T_2}{2},\ee
therefore we can use the saddle point method. The saddle (stationary phase) points are at
\be t_0={\sqrt{\alpha^2+1+\alpha^2T^2}-iT\over 1+T^2}\ ,\qquad  u_0=t_0^* .\ee
In the region of small $T$, after neglecting terms ${\cal O}(T^2)$,
\be \Big( \frac{t_0+u_0}{2}\Big)^2=1+\alpha^2=t_0u_0\ . \ee
Then  the integral yields
\be I(\mu, L)=2\pi^2(1+\alpha^2)^{L+1}\exp\Big[2\tan^{-1}(\alpha)-2\alpha\Big]{e^{-\mu\pi}\mu^{2(L+1)}\over k^2-L^2-\mu^2+i\epsilon} \, .\ee
After inserting this integral into Eq.(\ref{despro})  and using Stirling's formula for the limit  (\ref{limes}) of the Gamma function prefactor, we obtain
\be D_F(T_1,\Omega_1; T_2,\Omega_2)\approx 
\int_{-\infty}^{\infty}{dk\over 2\pi}e^{-ik(T_1-T_2)} \sum_{\L}{i\over k^2-L^2-\mu^2+i\epsilon}Y_{\bm{L}}(\Omega_1)Y_{\bm{L}}^*(\Omega_2)\, .\ee
 Since as mentioned before, we are considering short time intervals before and after the observer's $T=0$, we neglect terms ${\cal O}(T^2)$. Note that the above approximation is valid for the large $L$ part of the sum only. Then as shown below, for the points $\Omega_1=\bm{r_1}$ and $\Omega_2=\bm{r_2}$ near the North pole, parametrized as in  in Eq.(\ref{rid}), the sum can be converted into the integral:
\be \sum_{\L}{1\over k^2-L^2-\mu^2+i\epsilon}Y_{\bm{L}}(\Omega_1)Y_{\bm{L}}^*(\Omega_2)\approx 
\int\frac{d^3\bm{k}}{(2\pi)^3}{e^{i\bm{k}(\bm{r_1}-\bm{r_2})}\over k^2-\bm{k}^2-\mu^2+i\epsilon} \, .\label{dkk}\ee

In order to prove Eq.(\ref{dkk}), we consider
\be\int\frac{d^3\bm{k}}{(2\pi)^3}e^{i\bm{k}(\bm{r_1}-\bm{r_2})}f(|\bm{k}|)=\int\frac{L^2dLd^2\hat k}{(2\pi)^3}e^{i\bm{k}(\bm{r_1}-\bm{r_2})}f(L) \, .\ee
We express the plane waves as in Eq.(\ref{rhar}) and integrate over the directions $\hat k$ by using the orthogonality property of spherical harmonics. Then, since $L$ are large, we can replace the integral over $L$  by the sum:
\be\int\frac{d^3\bm{k}}{(2\pi)^3}e^{i\bm{k}(\bm{r_1}-\bm{r_2})}f(|\bm{k}|)\approx \sum_{\L}Y_{\bm{L}}(\Omega_1)Y_{\bm{L}}^*(\Omega_2)f(L)\ ,\ee
which completes the proof.

To summarize, we have a new integral representation (\ref{despro})  of the Feynman propagator. We showed that for very massive fields propagating in the neighborhood of the observer, the contribution of short wavelength modes coincides with the corresponding contributions to the flat space propagator. This new representation will be useful for discussing the infrared behaviour of scattering amplitudes.
\section{Scattering amplitudes}
\subsection{Infrared finiteness}
Dyson's formula  (\ref{dys}) is particularly useful for computing the  scattering amplitudes order by order in perturbation theory, by expanding the exponential in powers of the interaction terms. Each insertion of the   interaction term comes with an  integral over de Sitter volume, therefore it is important  to determine whether the integrals are convergent or not. The integrands involve products of wavefunctions and  propagators. The integration measure is
\be  d^4x \sqrt{-g}=\frac{1}{\cos^{4}t}\, dt\,d\Omega=(1+T^2)\, dT\, d\Omega\ .\label{meas}\ee
There are no convergence problems with the $d\Omega$ integrals over Eucliden (compact)  three-sphere; nevertheless, they are rather complicated because the integrands involve hyperspherical harmonics. There are potential problems, however, with the time integrals, at the endpoints $t=\pm \pi/2$ or equivalently, as $T\to \pm\infty$. The asymptotic behavior of associated Legendre functions can be determined by analytic continuation of hypergeometric functions. As $T\to\infty$,
\be
\widetilde{\P}^{-L-1}_{-\frac{1}{2}+i\mu}(i T) \approx\frac{1}{\sqrt{2\pi T}}\Big\{\frac{e^{\mu\pi\over 2}\Gamma(-i\mu)}{\Gamma (L+\frac{3}{2}-i\mu)}(2T)^{-i\mu}
+\frac{e^{-\mu\pi\over 2}\Gamma(i\mu)}{\Gamma (L+\frac{3}{2}+i\mu)}(2T)^{i\mu}\Big\}\ ,\label{as4}
\ee
up to a constant phase factor. This means that for real $\mu$, the wavefunctions (\ref{eq:phi4d}) behave as 
\be\phi(T,\Omega)_{\pm}\sim T^{-\frac{3}{2}}~~\makebox{as}~~ T\to\infty\ .\label{lart}\ee
It follows that the volume integrals (\ref{meas}) involving products of three or more wavefunctions are {\em absolutely} convergent. Note that in flat spacetime, similar integrals are only
{\em conditionally} convergent and usually distribution-valued. When $\mu$ is imaginary, $i\mu\equiv \mu_c\in \mathbb{R}$, as in the complementary series with $0<|\mu_c|<3/2$, infrared finitness holds for $|\mu_c|<1/2$. The critical values  $\mu=\pm i/2$, at which the three-point amplitude becomes logarithmically divergent $[\phi(T,\Omega)_{\pm}\sim T^{-1}]$, correspond to $m^2=9/4+\mu^2=2$ of the conformally coupled scalar. In the following, we will argue that the range $m^2<2$ of mass parameters should be excluded because the corresponding scalars behave as tachyons. This can be stated as: ``the physical mass squared is $m^2-2$.''

One  important comment/warning  is here in order. Some often quoted asymptotic expansions for the  Legendre functions of large order and/or degree are valid only in a limited range of  arguments. For example, the $L\to \infty$ asymptotics written in Eq.(\ref{pw}) are  valid only for  $t\in [-\pi/2+1/L, \pi/2-1/L]$ (or equivalently, for $|T|<L$). For large $T$, these approximate wavefunctions are suppressed as $T^{-1}$, therefore they fail to reproduce  the  stronger  $T^{-\frac{3}{2}}$ suppression, see Eq.(\ref{lart}). Fortunately, there exist so called ``uniform'' asymptotic expansions \cite{dunster}, valid on entire complex plane. These expansions involve Bessel functions and must be resorted to in cases when naive use of non-uniform expansions result in artificial  infrared ($T\to \infty$) divergences.
\subsection{Conservation laws}
Each insertion of the interaction term adds one Feynman vertex
integrated over the volume of dS, including the time direction and a three-sphere. Here, we focus on $d\Omega$ integrals over $S_3$. The integrands are product of hyperspherical harmonics (\ref{eq:Avery1}) originating from the wavefunctions and internal propagators.
They  factorize into the integrals of products of standard $S_2$ harmonics and the integrals of products of Gegenbauer polynomials. The integrals of spherical harmonics are familiar from quantum mechanics. Three-point vertices yield the simplest ones:
\begin{align}
\int_0^{\pi} d\theta \int_0^{2\pi} d\varphi \sin\theta & \, Y_{l_1n_1}(\theta,\varphi) Y_{l_2n_2}(\theta,\varphi) Y_{l_3n_3}(\theta,\varphi)  \nonumber\\
=~& \sqrt{\frac{(2l_1+1)(2l_2+1)(2l_3+1)}{4\pi}}  \Bigg( 
\begin{matrix}
l_1 & l_2 & l_3 \\
0 & 0& 0 
\end{matrix} \Bigg) \Bigg(
\begin{matrix}
l_1 & l_2 & l_3 \\
n_1 & n_2&  n_3
\end{matrix} \Bigg) \, , \label{eq:I3Y}
\end{align}
where 
$\scriptstyle
\Bigg(\begin{matrix}
l_1 & l_2 & l_3 \\
n_1 & n_2&  n_3
\end{matrix} \Bigg) \, 
$
is the well-known Wigner 3$j$-symbol. Recall that it is nonvanishing only if
\begin{align}
&n_1+n_2 +n_3=0 \, , \\
&l_1+l_2+l_3 \, \, \text{is an even integer.}\nonumber
\end{align}
Higher point vertices can be evaluated by using similar formulas. Then the above constraint reads
\begin{align}
&\sum_i n_i=0 \, , \label{ncons}\\
&\sum_il_i=0  \, \, \text{mod 2.}\nonumber
\end{align}
Hence one quantum number is conserved and the other one is conserved modulo 2. The integrals of Gegenbauer polynomials are given by more complicated expressions, but it is easy to show that they are nonvanishing only if
\be \sum_iL_i=0  \, \, \text{mod 2}\ ,\label{lcons}\ee
therefore we have another quantum number conserved modulo 2.

In the case of scattering amplitudes evaluated in the momentum basis in Minkowski spacetime,
similar arguments lead to four conserved quantities: the energy and three spatial momentum components.  Of course, four-momentum is conserved in all physical processes in flat spacetime, but in some bases, the symmetries of S-matrix are not so explicit and can be displayed only by using Ward identities. Here, we find  that dS amplitudes evaluated in the angular momentum basis are much less restricted by the conservation laws. The processes that are kinematically forbidden in flat spacetime are possible in dS. For example, as shown in the next section, a particle with mass $m$ can decay into two particles with the same mass $m$ - a process that is excluded by energy conservation in Minkowski spacetime.
\subsection{Decays}
In Part I, we studied several examples of scattering amplitudes in two-dimensional dS spacetime, in interacting scalar field theory with
\be H_I[\phi(x)]=\frac{\lambda}{3!}\phi^3(t,\Omega)\ .\label{hii}\ee
We showed that in the large angular momentum and mass limits,  in which the particles probe short spacetime intervals, de Sitter amplitudes agree with flat spacetime amplitudes. By following a similar line of arguments, it is not difficult to show that this property holds also in four dimensions. For that reason, we focus here on the processes that occur in deep infrared, when the particles have masses comparable to de Sitter scale (of order $10^{-33}$ eV$/c^2$ in our Universe) and/or carry low momenta. Here again, we consider scalars of the principal series, interacting with the Hamiltonian density (\ref{hii}). In this theory, the processes involving three external particles, like the decays of one  into two particles are simple to study because at the tree level, the corresponding amplitudes are given by the overlap integrals of three wavefunctions.

We consider the amplitude for the process in which particle number 1 decays into particles number 2 and 3, all in the principal series representation with the same $\mu$:
\be {}^{out}\langle\L_3\L_2|\L_1\rangle^{in}=i\lambda
\int d^4 x \sqrt{-g}\phi_{\bm{L}_3}(t,\Omega)_+\phi_{\bm{L}_2}(t,\Omega)_+ \phi_{\bm{L}_1}(t,\Omega)_- \, .
\ee
The overlap integral factorizes as follows:
\be
 {}^{out}\langle\L_3\L_2|\L_1\rangle^{in}=  i\lambda\,
\frac{|\prod_{k=1}^3\Gamma(L_k+\frac{3}{2}+i\mu)|}{(\sqrt{2})^3} I_{3P}I_{3Y}\ ,
\ee
with the spatial integral
\be I_{3Y}=\int d\Omega Y_{\bm{L}_3}(\Omega)Y_{\bm{L}_2}(\Omega)Y^*_{\bm{L}_1}(\Omega)\ee
which imposes the constraints (\ref{ncons}) and (\ref{lcons}) on the quantum numbers, in particular $L_1+L_2+L_3=0$ mod 2. The time integral is given by
\begin{align}
I_{3P}=&
\int_{-\infty}^\infty {dT \over \sqrt{1+T^2}}\widetilde{\bm{P}}^{-L_3-1}_{-\frac{1}{2}+i\mu}(iT) \widetilde{\bm{P}}^{-L_2-1}_{-\frac{1}{2}+i\mu}(iT)\widetilde{\bm{P}}^{-L_1-1}_{-\frac{1}{2}+i\mu}(-iT)\\[1.5mm] ~~= &
~2\,\Re\Big[ i^{L_3-L_2-L_1-1}\!\!\int_{0}^\infty \!\!{dT \over \sqrt{1+T^2}}{\bm{P}}^{-L_3-1}_{-\frac{1}{2}+i\mu}(iT) {\bm{P}}^{-L_2-1}_{-\frac{1}{2}+i\mu}(iT){\bm{P}}^{-L_1-1}_{-\frac{1}{2}+i\mu}(-iT) \Big],\nonumber 
\end{align}
where we used Eqs(\ref{meh1}) and (\ref{meh22}).

We begin with the  case of a particle with mass $m=3/2$, which is the minimum mass allowed by the principal series ($\mu=0$), at rest ($\L_1=0$), which decays into two particles with the same mass, also at rest ($\L_2=\L_3=0$). Such a process is forbidden by energy conservation in flat spacetime. It is allowed, however, in dS. In this case $I_{3Y}=(\sqrt{2}\pi)^{-1}$ and the time integral can be performed numerically:
\begin{equation}
I_{3P}(L_1=L_2=L_3=0; \mu=0) = 2.1268\pm 0.00005\ .
\end{equation}
As a result, after taking into account all factors,
\be
 {}^{out}\langle\bm{0}\bm{0}|\bm{0}\rangle^{in}\big|_{\mu=0}= \frac{i\lambda}{\sqrt{2}\pi} (0.5234\pm 0.00005)  \ .\ee

It is interesting to investigate how this amplitude changes as the mass  increases. The spatial part remains the same, while for large $\mu$, the Gamma function prefactor
\be |\Gamma(\frac{3}{2}+i\mu)|^3\sim e^{-{3\mu\pi\over 2}}\ .\ee
We can also use the asymptotic formulas for Mehler's functions with large $\mu$ \cite{dunster} and perform  time integrals numerically, As expected
when the masses increase, the amplitude is exponentially suppressed. This is shown in Fig.1, where we plot the ratio
\be \rho(\mu)=
\frac{\!\!{}^{out}\langle\bm{0}\bm{0}|\bm{0}\rangle^{in}\big|_{\mu}}{ ~~{}^{out}\langle\bm{0}\bm{0}|\bm{0}\rangle^{in}\big|_{\mu=0}}.\ee
\begin{figure}\centering\includegraphics[scale=0.4,page=1]{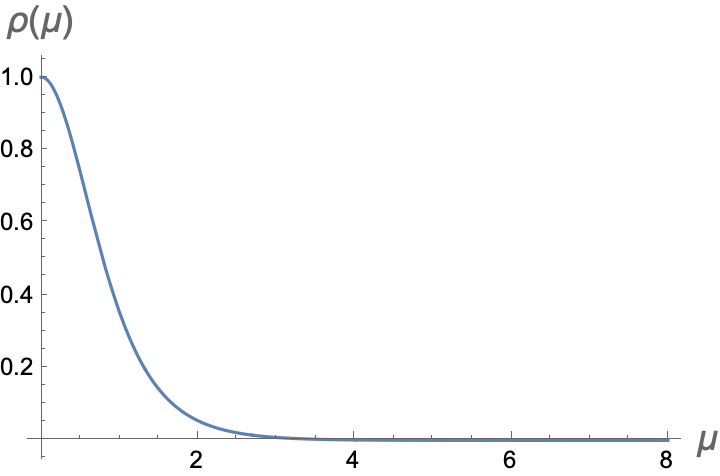}\vskip 0 cm\caption{Mass dependence of one to two particle decays. All particles with $\L=0$.}
\vskip -0 cm\end{figure}

We can also consider more exotic decays, when a particle at  rest decays into two particles with nonvanishing (angular) momentum:
$ {}^{out}\langle\L_2\L_3|\bm{0}\rangle^{in}$ with $\mu=0$ and  $L_2=L_3=L$. This process is also exponentially suppressed at large $L$, as shown in Figure 2, where we plot
\be \rho(L)=
\frac{\!\!{}^{out}\langle\bm{L}_2\bm{L}_3|\bm{0}\rangle^{in}\big|_{\mu=0}}{ ~{}^{out}\langle\bm{0}\bm{0}|\bm{0}\rangle^{in}\big|_{\mu=0}}~~\makebox{with} ~~L_2=L_3=L.\ee
\begin{figure}\centering\includegraphics[scale=0.4,page=1]{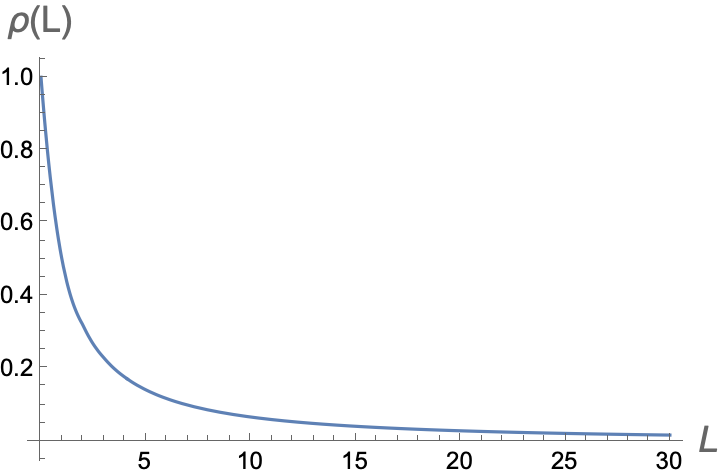}\vskip 0 cm\caption{Angular momentum dependence of one to two particle decays. All particles with $\mu=0$.}
\vskip -0 cm\end{figure}

To summarize, decays that are kinematically forbidden in flat spacetime are allowed in the deep infrared of dS, when the particles have masses of order of de Sitter scale $\hbar /c\ell $ and carry small (angular) momenta. Such processed are strongly suppressed, however, for larger masses and momenta, when the wavefunctions probe invariant distances much shorter than de Sitter radius $\ell$. 
\subsection{Particle production from ``nothing''}
A priori, the interaction term (\ref{hii}) allows annihilation of three scalars into the vacuum or  their creation from ``nothing.'' To be specific, we consider the latter process, in which three particles are created at rest, with vanishing  angular momenta. The corresponding amplitude reads
\be {}^{out}\langle\bm{0}\bm{0}\bm{0}|0\rangle^{in}=i\lambda
\int d^4 x \sqrt{-g}\big[\phi_{\bm{0}}(t,\Omega)_+\big]^3 \, .
\ee
In this case, the spatial integral is the same as for the  decays, while the time integral
\begin{align}
I_{3P}=&
\int_{-\infty}^\infty {dT \over \sqrt{1+T^2}}\widetilde{\bm{P}}^{-L_1-1}_{-\frac{1}{2}+i\mu}(iT) \widetilde{\bm{P}}^{-L_2-1}_{-\frac{1}{2}+i\mu}(iT)\widetilde{\bm{P}}^{-L_3-1}_{-\frac{1}{2}+i\mu}(iT)\label{noth}\\[1.5mm] ~~= &
~2\,\Re\Big[ i^{-L_3-L_2-L_1-3}\!\!\int_{0}^\infty \!\!{dT \over \sqrt{1+T^2}}{\bm{P}}^{-L_1-1}_{-\frac{1}{2}+i\mu}(iT) {\bm{P}}^{-L_2-1}_{-\frac{1}{2}+i\mu}(iT){\bm{P}}^{-L_3-1}_{-\frac{1}{2}+i\mu}(iT) \Big],\nonumber 
\end{align}
with $L_1=L_2=L_3=0$. We computed this integral numerically and found that it vanishes with high precision. We also computed some integrals with nonvanishing angular momenta and found that they are always zero as long as $L_2+L_2+L_3=0$ mod 2. Note that for $L_2+L_2+L_3=1$ mod 2, the spatial integrals are zero, therefore  particle creation from nothing is not possible in de Sitter spacetime. The vacuum is stable with respect to matter interactions. Unfortunately, there is no compact analytic formula available for the integrals  like (\ref{noth}). The stability of Bunch-Davis vacuum should follow, however, from de Sitter symmetry.
\subsection{Four-particle scattering and kinematic singularities} \label{sec:4-point}
In flat spacetime, the  scattering amplitudes exhibit kinematic singularities already at the tree level. They originate from Feynman diagrams with ``on-shell'' propagators, when the emmision or absorption  of a  particle propagating on an internal line becomes allowed by energy-momentum conservation. There, at $k_0^2=\bm{k}^2+m^2$ of the virtual particle, the propagator has  a simple pole. Such singularities are responsible for soft and collinear divergences \cite{taylor}. Do we encounter similar singularities in de Sitter amplitudes? To clarify this point, we will analyze the $\bm t$-channel exchange contribution to four-particle scattering in $\phi^3$ theory (\ref{hii}), as shown in Figure 3.

 \begin{figure}[htbp]\centering\includegraphics[scale=0.7,page=1]{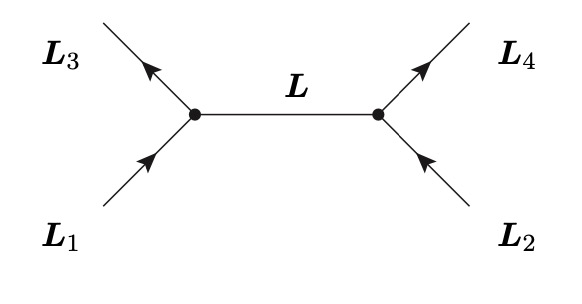}\vskip 0 cm\caption{$\bm{t}$-channel exchange diagram.}
\vskip -0 cm\end{figure}

\noindent It is given by
\begin{align} {}^{out}\langle\L_4\L_3|\L_2\L_1\rangle^{in}_{\bm t}=(i\lambda)^2&\int d^4 x_2 \sqrt{-g(x_2)}\int d^4 x_1 \sqrt{-g(x_1)}\, D_F(x_2,x_1)\nonumber
\\ &~~\times\phi_{\bm{L}_4}(t_2,\Omega_2)_+\phi_{\bm{L}_3}(t_1,\Omega_1)_+
\phi_{\bm{L}_2}(t_2,\Omega_2)_-\phi_{\bm{L}_1}(t_1,\Omega_1)_-\ .\label{four1}
\end{align}
To be  specific, we consider ``relativistic'' particles with  $L_i=l_i=n_i \gg \mu$. In this case, for a wide interval of $|T_i|<L_i$, the wavefunctions can be approximated by
\begin{align} \phi_{\bm{L}_i}(t_i,\Omega_i)_{+}\approx & ~\frac{1}{\sqrt{2L_i}}
\left(\frac{1- i T_i}{1+ iT_i} \right)^{\frac{L_i}{2}} \!Y_{\L_i}(\Omega_i) \, ,\nonumber\\  \phi_{\bm{L}_i}(t_i,\Omega_i)_{-}\approx & ~\frac{1}{\sqrt{2L_i}}
\left(\frac{1+ i T_i}{1 -iT_i} \right)^{\frac{L_i}{2}} \!Y_{\L_i}(\Omega_i)^*
\, .\end{align}
{}For a given angular momentum $\L$ flowing through the propagator, the contributions factorize as 
\be 
{}^{out}\langle\L_4\L_3|\L_2\L_1\rangle^{in}_{\bm t}=\frac{(i\lambda)^2}{|\Gamma(L+\frac{3}{2}-i\mu)|^2}(\prod_i\sqrt{2L_i})^{-1}I_{4P}I_{4Y}\ ,\ee
where the spatial integrals
\be I_{4Y}=\int d\Omega_2Y_{\L_4}(\Omega_2)Y_{\L_2}(\Omega_2)^*Y_{\L}^*(\Omega_2)
\int d\Omega_1Y_{\L_3}(\Omega_1)Y_{\L_1}(\Omega_1)^*Y_{\L}(\Omega_1)\ , \ee
and the time integrals
\begin{align}
I_{4P}=& \frac{i}{2\pi^2}\int_{-\infty}^{\infty}\!dT_1 dT_2 \left(\frac{1-i T_1}{1+iT_1} \right)^{\frac{\Delta_1}{2}}  \left(\frac{1-i T_2}{1+iT_2} \right)^{\frac{\Delta_2}{2}}  (1+T_1^2)^{L/2} (1+T_2^2)^{L/2}\label{tint}\\
&\times \int_0^\infty dt du (tu)^{L+1/2} (t+u) K_{-i\mu}(t) K_{-i\mu}(u) e^{-i(T_1+T_2)\frac{t-u}{2}} \int_{-\infty}^{+\infty} dk \frac{e^{-ik(T_1-T_2)}}{k^2-\left(\frac{t+u}{2} \right)^2+i\epsilon}\ , 
\nonumber\end{align}
where
\begin{equation}
\Delta_1=L_1-L_3, \quad \Delta_2=L_2-L_4 \, .
\end{equation}
Note that the spatial integrations impose constraints  on the angular momenta $\L$  propagating through the propagator.

First, consider scattering processes with zero angular momentum transfer: $\L_1=\L_3$ and $\L_2=\L_4$. 
In this case, it is easy to see that the dominant contribution comes from the $\L=\bm{0}$, exchange, with the integral $I_{4Y}=(2\pi^2)^{-1}$. In this case $\Delta_1=\Delta_2=L=0$ and the time integral  becomes
\begin{equation}
I_{4P}= 4 i\int_0^{\infty} dt \, t^2 K_{-i\mu}^2(t) \frac{1}{-t^2+i\epsilon} = -{i\pi^2\over \cosh (\pi\mu)} \, .
\end{equation}
In this way, we obtain
\begin{align}
{}^{out}\langle\L_2\L_1|\L_2\L_1\rangle^{in}_{\bm t}&=\frac{(i\lambda)^2}{|\Gamma(\frac{3}{2}-i\mu)|^2}(\prod_i\sqrt{2L_i})^{-1}I_{4P}I_{4Y} \label{zerot}\\[1mm]
&=i\lambda^2  (\prod_i\sqrt{2L_i})^{-1}  \frac{\pi}{\mu^2+\frac{1}{4}} I_{4Y}~=~
i\lambda^2  (\prod_i\sqrt{2L_i})^{-1}  \frac{\pi}{m^2-2} I_{4Y}\ ,\nonumber
\end{align}
where we used $m^2=\mu^2+9/4$.
In the large mass limit, when $\mu\approx m\gg 1$, after adjusting the kinematic factors from dS to flat spacetime as in Part I \cite{Taylor:2024vdc}, the above result agrees with the Minkowski amplitude.  Indeed, for the processes with zero momentum transfer the $\bm{t}$-channel diagrams contain the $1/m^2$ factor from the propagator.
Note that if one extrapolates  Eq.(\ref{zerot}) to the complementary series with  $\mu^2<0$, a singularity appears at $\mu^2=-1/4$, which corresponds to $m^2=2$ of a conformally coupled scalar. We see that  while dS geometry provides an infrared cutoff,  this cutoff becomes ``ineffective'' in a conformally coupled theory. For $m^2<2$, the $\bm{t}$-channel contribution changes the sign, which signals a tachyon. For that reason, the range $m^2<2$ of the mass parameter should be excluded. The physical mass squared of a scalar particle is  $m^2-2$.

Next, consider scattering processes with large momentum transfer, with $\Delta_1\gg 1$ and $\Delta_2\gg 1$, and focus on the $\bm{t}$-channel contribution of finite $L$, much smaller than $\Delta_1$ and $\Delta_2$.
In this case, it is convenient to rescale the integration variables as
 $T_1 \rightarrow \frac{T_1}{\Delta_1}$, $T_2\rightarrow \frac{T_2}{\Delta_2}$, so that in Eq.(\ref{tint})
\begin{equation}
\left(\frac{1-i T_1/\Delta_1}{1+iT_1/\Delta_1} \right)^{\frac{\Delta_1}{2}}  \left(\frac{1-i T_2/\Delta_2}{1+iT_2/\Delta_2} \right)^{\frac{\Delta_2}{2}} \approx e^{-i T_1} e^{-i T_2} \, ,
\end{equation}
and 
\be \big(1+\frac{T_1^2}{\Delta_1^2}\big)^{L/2} \big(1+\frac{T_2^2}{\Delta_2^2}\big)^{L/2}\approx 1\ .\ee
Then
\begin{align}
I_{4P} =2 i \int_0^\infty & dt du \int_{-\infty}^{\infty} dk \delta(k+ \frac{t-u}{2}+\Delta_L)\delta(-k+\frac{t-u}{2}+\Delta_R) \label{eq:DeltaLR} \\
&\times (tu)^{L+1/2} (t+u) K_{-i\mu}(t)K_{-i\mu}(u) \frac{1}{k^2-(\frac{t+u}{2})^2 +i\epsilon} \, . \nonumber
\end{align}
Let us further specify to a process in which the ``energy'' is conserved, with $L_1+L_2 = L_3+L_4$,
which corresponds to $\Delta_1=-\Delta_2\equiv\Delta$.
Then
\begin{align}
I_{4P} =4i \int_0^{\infty} du u^{2L+2} K_{-i\mu}^2(u) \frac{1}{-(u-\Delta)(u+\Delta) +i\epsilon} .
\end{align}
The integrand has a pole at $u=|\Delta|$. The integration is handled, however, by  the $+i\epsilon$ prescription which defines the integral as the principal value. There is no ``on-shell'' singularity for any value of $L$ or $\mu$.
Actually, for fixed $\mu$, when $ |\Delta|\gg L$, the integrand is suppressed by $K_{-i\mu}^2(u)\sim\pi (2u)^{-1}e^{-2u}$  at large $u$ and we can approximate\footnote{The integral contains also an imaginary part reflecting the instability of intermediate particle, but is it exponentially suppressed as $e^{-2|\Delta|}$ .}
\begin{align}
I_{4P} &\approx \frac{4 i}{\Delta^2} \int_0^\infty du u^{2L+2} K_{-i\mu}^2(u)  \nonumber\\
&=\frac{i}{\Delta^2} \frac{\sqrt{\pi}\Gamma(L+\frac{3}{2})|\Gamma(L+\frac{3}{2}-i\mu)|^2}{\Gamma(L+2)} \, ,
\end{align}
therefore
\be
{}^{out}\langle(\L_4\gg \L_2)(\L_3\gg\L_1)|\L_2\L_1\rangle^{in}_{\bm t}\approx -i\lambda^2  (\prod_i\sqrt{2L_i})^{-1} \frac{\sqrt{\pi}\Gamma(L+\frac{3}{2})}{\Gamma(L+2)\Delta^2}I_{4Y} \, .\ee

We conclude that ``on-shell'' singularities are absent in de Sitter spacetime.   Unlike in Minkowski spacetime, where virtual particles propagate inside the tree diagrams  with the momentum fixed by the Feynman rules, in de Sitter spacetime their (angular) momenta are less constrained.  The propagator singularities disappear after integrating over the (angular) momentum distributions because the +$i\epsilon$ prescription implies the principal value prescription for handling the propagator poles.

\section{Conclusions}
In this work, we discussed scalar quantum field theory in four-dimensional de Sitter spacetime. Our approach is centered on the observers who measure scattering amplitudes in their inertial reference frames. We discussed the properties of scattering amplitudes obtained by using the generalized Dyson's formula (\ref{dys}) derived in Part I. We discussed some basic elements of Feynman diagram computations. We derived a new integral representation of the Feynman propagator. While in Minkowski spacetime, the propagator propagates virtual particles with definite (off-shell) four-momentum, de Sitter propagator propagates virtual wavepackets with a wide spread of (angular) momenta. As a result, the kinematic singularities of the S-matrix (like the $\bm{s}$, $\bm{t}$, or $\bm{u}$ channel poles) due to virtual particles ``jumping on-shell'' are absent in de Sitter spacetime. Furthermore,  while the amplitudes evaluated in the momentum basis in flat spacetime are distribution-valued (they contain momentum-conserving delta functions due to translational invariance), in de Sitter spacetime the amplitudes are well-defined functions of quantum numbers. The infrared divergences, which are normally  related to the infinite spacetime volume, are absent in de Sitter spacetime: spatial dimensions are compact while the wavefunctions are suppressed in the noncompact time direction.

In Part I, we focused on ``ultraviolet'' processes involving quantum waves with large frequencies and short wavelengths and showed that their scattering amplitudes, which probe short spacetime intervals, agree with Minkowski amplitudes. In this work, we discussed deep infrared processes involving cosmologically light particles with low (angular) momenta. We focussed on the processes that are normally forbidden in flat spacetime by the energy-momentum conservation laws: one particle decays into two particles of the same mass and on the ``all out'' amplitudes for multiparticle production from the vacuum. In interacting scalar free theory, there are no stable particles, although the decay rates are exponentially suppressed for heavier particles.   On the other hand, ``all out'' correlators vanish, which ensures  the stability of Bunch-Davies vacuum with respect to the matter interactions. We also discussed similar four-particle amplitudes and found one example of a kinematic singularity due to the propagation of a conformally coupled scalar.
We expect that such singularities appear in conformally invariant theories only, which are insensitive to the presence of the cosmological cutoff.

This work was limited to interacting scalar field theory. The ultimate goal, however, is to include spinning particles, gauge interactions, and gravitational interactions due to the fluctuating geometry. Work in this direction is in progress.

It is clear that the computations of de Sitter scattering amplitudes are technically more complicated than the computations of Minkowski amplitudes. They involve integrals of associated Legendre functions and hyperspherical harmonics. We hope that a simple mathematical, more abstract framework can be developed for such computations.

\section*{Acknowledgments}
This work was supported in part by NSF PHY-2209903, the Simons Collaboration on Celestial Holography, and
Polish National Agency for Academic Exchange under the NAWA Chair programme.
It was also supported by the MAESTRO grant no.\ 2024/54/A/ST2/00009
funded by the National Science Centre, Poland.
Any opinions, findings, and conclusions or
recommendations expressed in this material are those of the authors and do not necessarily
reflect the views of the National Science Foundation.

\appendix 
\section{Associated Legendre functions and Gegenbauer polynomials} \label{app:A}
Most of the special functions utilized in the main text were already discussed in the Appendix of Part I. Here, we list more formulas relevant to the present work. 

The associated Legendre functions are solutions of the differential equation
\begin{equation}
(1-z^2) \frac{d^2 u}{d z^2} -2 z \frac{du}{d z} +\left[ \nu(\nu+1) - \frac{\mu^2}{1-z^2}\right] u = 0 \, ,
\end{equation}
where the parameters $\mu$ and $\nu$ are referred to as the order and degree, respectively. In general, they are complex numbers. 
They are defined on the complex plane with a branch cut running from  from $-{}\infty$ to $+{}1$.
There are two types of linearly independent associated Legendre functions: $\P$ and $\Q$.  They are usually defined through hypergeometric functions $_2F_1$:
\begin{equation}
\P^{\mu}_{\nu}(z) = \frac{1}{\Gamma(1-\mu)} \left( \frac{z+1}{z-1}\right)^{\mu/2} \, _2F_1\left( -\nu, \nu+1; 1-\mu; \frac{1-z}{2} \right) \, ,
\end{equation}
\begin{equation}
\Q^{\mu}_{\nu}(z) = \frac{e^{\mu\pi i}\Gamma(\nu+\mu+1)\Gamma(\frac{1}{2})}{2^{\nu+1}\Gamma\left(\nu+\frac{3}{2}\right)} (z^2-1)^{\mu/2} z^{-\nu-\mu-1} \,_2F_1\left(\frac{\nu+\mu+2}{2}, \frac{\nu+\mu+1}{2}; \nu+\frac{3}{2}; \frac{1}{z^2} \right) \, .
\end{equation}
$\P$ and $\Q$ are also called associated Legendre functions of the first and second kind, respectively.

The associated Legendre function $\Q$ above or below the cut is related to the 
Ferrers' functions,
\begin{equation}
e^{-\mu \pi i} \Q^{\mu}_{\nu}(x\pm i\epsilon) = e^{\pm \frac{1}{2}\mu\pi i} \left[ Q^{\mu}_{\nu}(x) \mp i \frac{\pi}{2} P^{\mu}_{\nu}(x)\right] \, . \label{eq:defQcut}
\end{equation}
The reflection formulas for the associated Legendre functions are
\begin{equation}
\P^{\mu}_{\nu}(-z) = e^{\mp i\pi \nu} \P^{\mu}_{\nu}(z) -\frac{2}{\pi} \sin[\pi(\nu+\mu)] e^{-\mu\pi i}\Q^{\mu}_{\nu}(z) \, , 
\end{equation}
\begin{equation}
\Q^{\mu}_{\nu}(-z) = - e^{\pm i \pi \nu} \Q^{\mu}_{\nu}(z) \, , \label{eq:refformula}
\end{equation}
where the upper or lower sign needs to be taken according to Im$z>0$ or Im$z<0$. 

In the main text, we used Whipple's formula
\begin{align}
\Q^{\mu}_{\nu}(z) &= e^{i\pi \mu} \left(\frac{1}{2}\pi\right)^{\frac{1}{2}} \Gamma(\nu+\mu+1) (z^2-1)^{-\frac{1}{4}} \P^{-\nu-\frac{1}{2}}_{-\mu-\frac{1}{2}}\left[ z (z^2-1)^{-\frac{1}{2}}\right] \, , \label{eq:Whipple}
\end{align}
which is valid for $\Re z>0$.

The following relation between Legendre P and hypergeometric functions was used to obtain the asymptotic formula (\ref{as4}):
\begin{align}
\P^{\mu}_{\nu}(z) =&  \frac{2^{-\nu-1} \pi^{-\frac{1}{2}} \Gamma(-\frac{1}{2}-\nu) z^{-\nu+\mu-1}(z^2-1)^{-\frac{1}{2}\mu}}{\Gamma(-\nu-\mu)} \, _2F_1\left({\frac{1}{2}+\frac{1}{2}\nu-\frac{1}{2}\mu, 1+\frac{1}{2}\nu -\frac{1}{2} \mu \atop \nu +\frac{3}{2}} \, ; z^{-2} \right) \nonumber\\
&+ \frac{2^{\nu} \pi^{-\frac{1}{2}} \Gamma(\frac{1}{2}+\nu) z^{\nu+\mu} (z^2-1)^{-\frac{1}{2} \mu}}{\Gamma(1+\nu-\mu)} \, _2F_1\left({-\frac{1}{2}\nu-\frac{1}{2}\mu, \frac{1}{2}-\frac{1}{2}\nu -\frac{1}{2} \mu \atop 1-\nu } \, ; z^{-2} \right) \, .
\end{align}

Next, we discuss the Gegenbauer polynomials which are used for constructing hyperspherical harmonics. The Gegenbauer polynomials $C_n^\lambda(t)$ of degree $n$ are the coefficients of $x^n$ in the series expansion
\begin{equation}
(1-2t x+x^2)^{-\lambda} = \sum_{n=0}^\infty C_n^\lambda(t) x^n \, .
\end{equation}
The general expression for the polynomials is 
\begin{equation}
C_n^\lambda(t) = \sum_{k=0}^{\lfloor n/2\rfloor} (-1)^k \frac{\Gamma(n-k+\lambda)}{\Gamma(\lambda) k!(n-2k)!} (2t)^{n-2k} \, ,
\end{equation}
where $\lfloor x \rfloor$ is the floor function that takes the integer part of a positive number.
They can be expressed in terms of hypergeometric functions in the following way:
\begin{equation}
C_n^\lambda(t) = \frac{\Gamma(2\lambda+n)}{\Gamma(n+1)\Gamma(2\lambda)} \, _2F_1\left( 2\lambda+n, -n; \lambda+\frac{1}{2}; \frac{1-t}{2}\right) \, .
\end{equation}
There are formulas for even and odd orders respectively,
\begin{equation}
C_{2n}^\lambda(t) = \frac{(-1)^n}{(\lambda+n)B(\lambda,n+1)} \, _2F_1\left(-n,n+\lambda; \frac{1}{2}; t^2\right) \, . \label{eq:C2n}
\end{equation} 
\begin{equation}
C_{2n+1}^\lambda(t) = \frac{(-1)^n 2t}{B(\lambda,n+1)} \, _2F_1\left( -n, n+\lambda+1; \frac{3}{2}; t^2\right) \, . \label{eq:C2n+1}
\end{equation} 
The Gegenbauer polynomials are orthogonal in the following sense:
\begin{align}
&\int_{-1}^1 (1-x^2)^{\nu-\frac{1}{2}} C_m^{\nu}(x) C_m^\nu(x) dx = 0  \quad \quad m\neq n \, ,\\
&\int_{-1}^1 (1-x^2)^{\nu-\frac{1}{2}} [C_n^\nu(x)]^2 dx = \frac{\pi 2^{1-2\nu} \Gamma(2\nu+n)}{n! (n+\nu) \Gamma(\nu)^2} \, .
\end{align}

There are two formulas for the integrals over Gegenbauer polynomials, which can be used for computing the overlaps of hyperspherical harmonics.
We are interested in the integral of the following form,
\begin{equation}
\int_{-1}^1 dy (1-y^2)^{\frac{l+1}{2} +\frac{b}{2}} y^{a+n} C_{L-l}^{1+l}(y) \, . \label{eq:IntCL-l}
\end{equation}
This  integral depends on whether  $a+n$ is even or odd. Due to the  properties of Gegenbauer polynomials, when $a+n$ is odd, $L-l$ must be odd. When $a+n$ is even, $L-l$ must be even.

1) When $a+n$ and $L-l$ are even, by using (\ref{eq:C2n}), we find 
\begin{align}
&\int_{-1}^1 dy (1-y^2)^{\frac{l+1}{2}+\frac{b}{2}} y^{a+n} C_{L-l}^{1+l}(y)  \nonumber\\
=& \frac{(-1)^{\frac{L-l}{2}}}{\left( \frac{L+l+2}{2}\right)B\left(1+l, \frac{L-l+2}{2}\right)}  \Gamma\left(1+\frac{b}{2}+ \frac{1+l}{2} \right) \Gamma\left(\frac{1}{2}+\frac{a+n}{2}\right) \frac{1}{ \Gamma\left(2+\frac{b}{2}+\frac{l}{2}+\frac{a+n}{2} \right)} \nonumber\\
&\times \, _3F_2\left( {\frac{1}{2}(l-L), \frac{1}{2}(2+l+L), \frac{1}{2}+ \frac{a+n}{2} \atop \frac{1}{2}, 2+\frac{b}{2} +\frac{l}{2}+\frac{a+n}{2}} ; 1\right) \,  . \label{eq:int1C}
\end{align}

2) When $a+n$ and $L-l$ are odd, using (\ref{eq:C2n+1}), we find
\begin{align}
&\int_{-1}^1 dy (1-y^2)^{\frac{l+1}{2}+\frac{b}{2}} y^{a+n} C_{L-l}^{1+l}(y)  \nonumber\\
=& \frac{(-1)^{\frac{L-l-1}{2}}}{B\left(1+l, \frac{L-l+1}{2}\right)}  \Gamma\left(1+\frac{b}{2}+ \frac{1+l}{2} \right) \Gamma\left(1+\frac{a+n}{2}\right) \frac{2}{ \Gamma\left(2+\frac{b}{2}+\frac{1+l}{2}+\frac{a+n}{2} \right)} \nonumber\\
&\times \, _3F_2\left( {\frac{1}{2}(1+l-L), \frac{1}{2}(3+l+L), 1+ \frac{a+n}{2} \atop \frac{3}{2}, 2+\frac{b}{2} +\frac{1+l}{2}+\frac{a+n}{2}} ; 1\right) \,  .
\end{align}

Eq.(\ref{eq:int1C}) can be used to compute the spatial part integral of the four-point exchange diagram that we considered in Sec.\ref{sec:4-point} when the external $L_i=l_i$,
\begin{align}
I_{4Y}&\sim \int_{-1}^{1}dy \sqrt{1-y^2}^{1+l_1+l_3+l} C_{L-l}^{1+l}(y) \int_{-1}^1 dy \sqrt{1-y^2}^{1+l_2+l_4+l} C_{L-l}^{1+l}(y) \nonumber\\[1mm]
&= \frac{\pi}{\left( \frac{L+l+2}{2}\right)^2B\left(1+l, \frac{L-l+2}{2}\right)^2} \frac{\Gamma(1+\frac{l_1+l_3+l+1}{2})}{\Gamma(2+\frac{l_1+l_3+l}{2})}\, _3F_2\left( {\frac{1}{2}(l-L), \frac{1}{2}(2+l+L), \frac{1}{2} \atop \frac{1}{2}, 2+\frac{l_1+l_3}{2} +\frac{l}{2}} ; 1\right) \nonumber\\[2mm]
&~~~~\times\frac{\Gamma(1+\frac{l_2+l_4+l+1}{2})}{\Gamma(2+\frac{l_2+l_4+l}{2})}\, _3F_2\left( {\frac{1}{2}(l-L), \frac{1}{2}(2+l+L), \frac{1}{2} \atop \frac{1}{2}, 2+\frac{l_2+l_4}{2} +\frac{l}{2}} ; 1\right) \, .
\end{align}

\end{document}